\documentclass[aip,nofootinbib]{revtex4}
\usepackage{graphicx}
\usepackage{dcolumn}
\usepackage{bm}

\begin{document}

\preprint{  }

\title{On relativistic motion of a pair of particles having opposite signs of masses}

\author{Pavel Ivanov}
\email{pbi20@cam.ac.uk}
\affiliation{%
Astro Space Centre of PN Lebedev Physical Institute,
84/32 Profsoyuznaya Street, Moscow, 117810, Russia}%


\begin{abstract}

In this methodological note we consider, in a weak-field limit, a
relativistic linear motion of two particles with opposite signs of
masses and a small difference between their absolute values
$m_{1,2}=\pm (\mu\pm \Delta \mu) $, $\mu > 0$, $|\Delta \mu | \ll
\mu$. In 1957 H. Bondi  showed both in framework of Newtonian
analysis and in General Relativity that when the relative motion
of particles is absent such a pair can be accelerated
indefinitely. We generalise results of his paper to account for a
small nonzero difference between velocities of the particles.

Assuming that the weak-field limit holds and the dynamical system
is conservative an elementary treatment of the problem based on
the laws of energy and momentum conservation shows that the system
can be accelerated indefinitely, or attain very large asymptotic
values of the Lorentz factor $\gamma$. The system experiences
indefinite acceleration when its  energy-momentum vector is null
and the mass difference $\Delta \mu \le 0$. When modulus of the
square of the norm of the energy-momentum vector, $|N^2|$, is
sufficiently small the system can be accelerated to very large
$\gamma \propto |N^2|^{-1}$.

It is stressed that when only leading terms in the ratio of a
characteristic gravitational radius to the distance between the
particles are retained our elementary analysis leads to equations
of motion equivalent to those derived from relativistic weak-field
equations of motion of Havas and Goldberg 1962.

Thus, in the weak-field approximation, it is possible to bring the
system to the state with extremely high values of $\gamma $. The
positive energy carried by the particle with positive mass may be
conveyed to other physical bodies say, by intercepting this
particle with a target. Suppose that there is a process of
production of such pairs and the particles with positive mass are
intercepted while the negative mass particles are expelled from
the region of space occupied by physical bodies of interest. This
scheme could provide a persistent transfer of positive energy to
the bodies, which may be classified as a 'Perpetuum Motion of
Third Kind'.

Additionally, we critically evaluate some recent claims on the
problem.

\end{abstract}

\maketitle

\section{Introduction}

Bondi 1957 [1] pointed out that in the Newtonian approximation two
particles with opposite signs of masses at rest with respect to
each other accelerate indefinitely in an inertial frame. This
process is allowed by the laws of conservation  since the kinetic
energy and angular momentum of such a system are conserved being
exactly zero while the potential energy depends only on relative
distance between the particles. In the same paper he generalised
this result by finding an appropriate static accelerated solution
in General Relativity and discovered that a uniformly accelerated
pair of particles with opposite signs of masses must have a mass
difference determined by the fact that constant in time particle
accelerations  must be different to keep them static with respect
to each other.

It is trivial to show that in the Newtonian approximation (see the
next Section ) when the two particles with opposite signs of
masses have a relative velocity, its value is approximately
conserved. As a result, the acceleration period is finite and the
pair as a whole being initially at rest gains a finite value of
velocity. We also show that when initial relative velocity of the
particles is sufficiently small the pair can be accelerated to a
relativistic speed.

In the next Section we consider the problem in the relativistic
setting and generalise Bondi's analysis considering pairs of
particles with opposite masses and small difference between their
absolute values: $m_{1,2}=\pm (\mu\pm \Delta \mu) $, $\mu
> 0$, $|\Delta \mu | \ll \mu$  having an initial relative velocity
$v_{in}$ in a fixed lab frame where the pair as a whole is
initially at rest. We assume that gravitational interaction is
weak and, therefore, $G\mu/(c^2 D_{in})\ll 1$, where $D_{in}$ is
an initial separation distance between the particles. Also, for
simplicity, in the relativistic treatment, it is assumed that the
orbital angular momentum of the system is equal to zero and the
motion is linear.

We analyse this situation by elementary means. The equations of
motion are obtained from the laws of energy and momentum
conservation. It is assumed that energy and momentum of the system
in a Lorentz frame instantaneously comoving with the motion of the
pair are given by the Newtonian expressions and that they form
time and spacial components of a local four-vector. Then this
energy-momentum vector is projected onto the lab frame. Since
energy and momentum in the lab frame are conserved under the
assumption that gravitational radiation from the system is
insignificant we get two equations of first order in time fully
describing dynamics of the system. We also show how to derive an
equivalent pair of second order  equations considering the
Newton's law of gravity in a frame accelerating with the
particles.

It is shown that the pair as a whole always have a positive
acceleration with its asymptotic value  being either zero or a
nonzero constant depending on initial conditions. The relative
distance between the particles can either have a turning point or
increase monotonically. The system accelerates indefinitely when
the mass difference $\Delta \mu \le 0$ and the norm of the
energy-momentum vector $N=\sqrt{(2\Delta \mu c^2 +{G\mu^{2}\over
D_{in}})^{2}-\mu^2v_{in}^2}=0$, and, accordingly, the
energy-momentum vector is null. In this case the relative distance
increases monotonically. When $N^2$ is sufficiently small, for the
initial conditions corresponding to the monotonic behaviour of the
relative distance the acceleration period is finite but the
asymptotic value of the Lorentz gamma factor is large being
proportional to $|N^{2}|^{-1}$.

Such pairs can play a role in realisation of a hypothetical
effect, which we called 'Perpetuum Motion of Third Kind' [2],
hereafter PMT. In its most general formulation this effect is a
possibility of a persistent energy transfer from a subsystem
having negative energy to a subsystem with positive energy, in
classical theories where negative energy subsystems are possible.
Indeed, the positive mass particle can, in principle, be used to
transfer positive energy to other physical bodies after the pair
has been accelerated to high values of the Lorentz factor.
Iterating this process as many times as we need we can extract as
much positive energy as we wish. Note, however, that it is not the
only 'working model' of PMT, and that, in principal, in order to
make PMT we need neither systems with negative rest mass nor
gravitational interactions. As is shown in [2] it suffices to have
a medium violating the weak energy condition with certain
additional properties and mere hydrodynamical interaction 'to
construct a PMT'.

Additionally, we comment on several statements of paper [3], where
the Kepler problem for a binary with opposite signs of masses has
been considered which may, in our opinion, lead to
misunderstanding of the problem.

\section{Newtonian treatment of the problem}

At first let us consider the problem in the Newtonian
approximation where mutual gravitational accelerations acting on
the particles of masses $m_1$ and $m_2$ are given by the
conventional expressions:
\begin{equation}
\ddot {\bf r}_1=-{Gm_2\over |{\bf D}|^3}{\bf D},  \quad \ddot {\bf
r}_2={Gm_1\over |{\bf D}|^3}{\bf D}, \label{e1}
\end{equation}
where ${\bf r}_i$ are position vectors of particles with indices
$i=1,2$ and ${\bf D}={\bf r}_{1}-{\bf r}_2$. Setting $\mu \equiv
Gm_1=-Gm_2$ we obtain from (\ref{e1})
\begin{equation}
\dot {\bf V}= {\mu \over |{\bf D}|^3}{\bf D}, \quad \dot {\bf v} =
0, \label{e2}
\end{equation}
where ${\bf V}\equiv {1\over 2}(\dot {\bf r}_1+ \dot {\bf r}_2)$
and ${\bf v}\equiv \dot {\bf D}$. From equation (\ref {e2}) it
follows that when ${\bf v}=0$ at some moment of time it remains
zero in the course of evolution of the system. Thus, in this case
the relative separation distance ${\bf D} $ does not change during
the evolution and the system ever accelerates as a whole, with the
acceleration vector
\begin{equation}
{\bf a}\equiv \dot {\bf V}= {\mu \over |{\bf D}|^3}{\bf D}
\label{e3}
\end{equation}
being constant. The laws of conservation are nonetheless respected
since the kinetic energy and momentum of the system are precisely
zero while the potential energy depends only on the relative
separation distance\footnote{Note that it is easy to show that the
same motion can be realised in a system containing $N$ particles
provided that the total mass of the system $M=\sum_{i=1,n} m_i =0
$ and positions of the particles are chosen in a special way. Say,
for a system containing three particles their relative positions
must form an equilateral triangle.}.

When ${\bf v}(t=0)\equiv {\bf v}_{in}\ne 0$ the absolute value of
the relative distance changes with time. Accordingly, the absolute
value of the acceleration changes as well and eventually decays
provided that $({\bf D}_{in} \cdot {\bf v}_{in}) \ne - |{\bf
v}_{in}||{\bf D}_{in}|$\footnote{Clearly, when $({\bf D}_{in}
\cdot {\bf v}_{in}) = - |{\bf v}_{in}||{\bf D}_{in}|$ the
particles collide.}, where ${\bf D}_{in}\equiv {\bf D}(t=0)$. We
have
\begin{equation} {\bf D}={\bf v}_{in}t + {\bf
D}_{in}, \label{e4}
\end{equation}
and, thus, integrating equation (\ref{e3}) we obtain
\begin{equation}
|{\bf V}(t)|={1\over \sqrt{1-\alpha^2}}{\mu \over D_{in} v_{in}}
\sqrt {2(1-{\epsilon^{1/2}+\alpha \tau\over \Delta})}, \label{e5}
\end{equation}
where we use the dimensionless time $\tau =\sqrt{{\mu \over
D_{in}^3}}t$, $D_{in}=|{\bf D}_{in}|$, $v_{in}=|{\bf v}_{in}|$,
$\epsilon ={\mu \over D_{in} v_{in}^2}$, $\alpha =({\bf
v}_{in}\cdot {\bf D}_{in})/(v_{in}D_{in})$, and $\Delta =\sqrt
{\epsilon +\tau^2+2\alpha \epsilon^{1/2}\tau}$.

Note that when the system moves along a straight line with
increasing value of $|{\bf D}|$, and, accordingly, $\alpha =1 $
equation (\ref{e5}) yields
\begin{equation}
|{\bf V}(t)|={\mu \over D_{in} v_{in}}{\tau \over
\epsilon^{1/2}+\tau}. \label{e6}
\end{equation}

In the limit $\tau \rightarrow \infty $ we get from equations
(\ref{e5}) and (\ref{e6})
\begin{equation}
V_{\infty}\equiv |{\bf V}(\tau \rightarrow \infty )|=\sqrt
{{2\over 1 +\alpha }}{\mu \over D_{in} v_{in}}. \label{ee7}
\end{equation}
It follows from (\ref{ee7}) that when
\begin{equation}
v_{in} < v_{crit}= \sqrt {{2\over 1 +\alpha }}{\mu \over D_{in} c}
\label{e7}, \label{e8}
\end{equation}
the asymptotic value of velocity of the system, $V_{\infty}$,
formally exceeds the speed of light, $c$. Clearly, a relativistic
approach to the problem is to be used in this situation.

\section{Relativistic treatment}

\subsection{Derivation of dynamical equations}

In order to keep our study as simple as possible let us consider
in the relativistic case only the motion along a straight line
with increasing value of ${\bf D}$ ($\alpha = 1$). Additionally,
in this Section we use the natural units setting the speed of
light and the gravitational constant to unity. However, unlike the
Newtonian case, here we would like to consider particles having a
small mass difference: $m_{1,2}=\mu \pm \Delta \mu$, where it is
assumed below that $\mu
> 0$ and $|\Delta \mu | \ll \mu$.

It is useful to introduce two local frames and the respective
coordinate systems 1) a fixed lab frame with global Lorentzian
coordinates $(x,t)$ and 2) a local Lorentzian  frame
instantaneously comoving with the motion of the point
$R(t)={1\over 2}(x_{1}(t)+x_{2}(t))$, where $x_1(t)$ and
$x_{2}(t)$ are positions of the particles in the lab frame, with
associated Lorentzian coordinates $(x^{com},t^{com})$.  Is is
assumed that at some particular moment of time $t=t_*$ coordinates
of the event $(t_*, R(t=t_*))$ in the comoving coordinate system
are equal to $(\tau, 0)$, where $\tau$ is the proper time
associated with the world line $(t, R(t))$. Hereafter, the world
line $(t, R(t))$ is referred to as "the reference world line".

When $t^{com}=\tau$ the positions of particles are given by
$x^{com}_{1,2}(t^{com})$, their velocities are
$v^{com}_{1,2}={d\over dt^{com}}x^{com}_{1,2}$. Let us also
introduce the relative position and velocity in the comoving
coordinate system $D=x_1^{com}-x_2^{com}$, $v^{com}={d\over
dt^{com}}D$. Without loss of generality we assume hereafter
$D^{com} > 0$. When the relative separation remains sufficiently
small along the reference world line we  have approximately
$x^{com}_{2}=-x^{com}_{1}$.

In the global coordinates at the time slice $t=t_*$ the velocity
of motion of the system as a whole is given by $V={1\over
2}({d\over dt}x_1+{d\over dt}x_2)(t=t_*)$ while the relative
position and velocity of the relative motion are
$D_{lab}=x_1(t_*)-x_2(t_*)$ and $v={d\over dt}D$.

Introducing the Lorentz gamma factor $\gamma ={1\over
\sqrt{1-V^{2}}}$ associated with the reference world line we may
write in the limit of small separations
\begin{equation}
D(t^{com})=\gamma D_{lab}, \quad {dt\over dt^{com}}=\gamma ,
\label{e9}
\end{equation}
and, accordingly,
\begin{equation}
v^{com}=\gamma{d\over dt}(\gamma D_{lab})=\gamma^{2}v+\gamma
{d\gamma \over dt}D_{lab}. \label{e10}
\end{equation}

Supposing  below that, on one hand, the relative distance $D \gg
\mu$, and, therefore, a weak-field approximation holds and, on the
other hand, it is not too large for the local Lorentzian
coordinates to be adequate and, respectively, equations
(\ref{e9}-\ref{e10}) to be valid we can use the Newtonian
expression for the energy, $E_c$, and momentum, $P_c$, of the
system in the comoving frame at the time $t^{com}=\tau$
\begin{equation}
E_c=2\Delta \mu +{\mu^2\over D}, \quad P_c=\mu \dot D, \label{e11}
\end{equation}
where dot stands for differentiation w.r.t. the proper time
$\tau$.

In the same limit $E_c$ and $P_c$ represent time and spacial
components of a local four vector, and, therefore, they values in
the lab frame, $E$ and $P$, respectively, can be obtained from
(\ref{e11}) by the standard Lorentz transformation. We have
\begin{equation}
E=\gamma (2\Delta \mu +{\mu^2\over D} + V\mu \dot D), \quad
P=\gamma ( \mu \dot D + V(2\Delta \mu +{\mu^2\over D})),
\label{e12}
\end{equation}
where it is assumed that the velocity of the systems as a whole,
$V$, is a function of the proper time $\tau $. Since energy and
momentum in the lab frame are obviously conserved, equations
(\ref{e12}) fully describe the dynamics of our system. They should
be solved subject to the condition that the system is initially at
rest with respect to the lab frame: when $\tau =0$ we have $V=0$
and \begin{equation} E=E_{in}=2\Delta \mu +{\mu^2\over D_{in}},
\quad P=P_{in}= \mu v_{in}, \label{e12a} \end{equation}
 where $D_{in}$ and $v_{in}$ are initial separation distance and relative
velocity, respectively. It is assumed below that $v_{in} > 0$.

Although our derivation of dynamical equations (\ref{e12}) may
look somewhat heuristic it is worth mentioning that when terms
next to the leading order in $\mu $ are discarded they can be
derived from the precise weak-field equations of reference [4] in
the limit of small separations and $|\Delta \mu | \ll \mu$ .

It is convenient to transform equations (\ref{e12}) to another
form using their linear combination $E-VP$ and calculating square
of the norm of the energy-momentum vector, $N^{2}=E^2-P^2$. We get
\begin{equation}
E-VP=\gamma^{-1}(2\Delta \mu +{\mu^2\over D}) \label{e13}
\end{equation}
and
\begin{equation}
N^{2}= (2\Delta \mu +{\mu^2\over D})^{2}-\mu^{2}(\dot D)^2.
\label{e14}
\end{equation}
We also obviously have $N^{2}=(2\Delta \mu +{\mu^2\over
D_{in}})^{2}-\mu^{2}v_{in}^2$. Note that contrary to the usual
situation the energy-momentum vector can be null, time-like or
space-like, depending on initial conditions.

Equations (\ref{e12}) are first order integrals of two second
order in time dynamical equations. One of these  equations can be
obtained from (\ref{e14}) by differentiating this equation over
$\tau $ with the result
\begin{equation}
\ddot D= -{2\Delta \mu \over D^{2}}-{\mu^2\over D^3}, \label{en1}
\end{equation}
and the second one by differentiating either of equations
(\ref{e12}) and using (\ref{en1}):
\begin{equation}
\gamma^{2}\dot V={\mu \over D^2} \label{en2}
\end{equation}

Equations (\ref{en1}) and (\ref{en2}) can be obtained from other
independent qualitative arguments. The derivation of the second
order dynamical equations, which relate dynamical variables with
different values of time coordinates is not, however, convenient
in the local Lorentzian coordinates introduced above since these
coordinates are defined with respect to some particular event on
the reference world line and, therefore, the definition is
different for different events along this world line. It is much
more convenient to use a coordinate system, where the proper time
$\tau $ plays the role of coordinate time. For that let us
consider another, the so-called local Fermi-Walker coordinate
system $(\tau, y)$, see e.g. [5], where the proper time $\tau $ is
the coordinate time and the unit vector in the spacial direction
$y$ is always perpendicular to the four velocity along the
reference world line. The coordinates of the reference world line
in this coordinate system are simply $(\tau, 0)$.

At the time slice $t^{com}=\tau $ the local Lorentz coordinates
and the Fermi-Walker coordinates coincide:
$x^{com}_{1,2}=y_{1,2}$, but the Fermi-Walker coordinate system is
accelerating with respect to the local Lorentz coordinate system
with an acceleration $g(y)$. Clearly, $g(y=0)$ must coincide with
modulus of four acceleration of the reference world line with
respect to the lab frame. The equations of motion in the
Fermi-Walker coordinates are assumed to be determined by the
Newton's law (\ref{e1}) with added acceleration term $-g$, which
accounts for the fact that this system is not inertial:
\begin{equation}
\ddot y_{1,2}={\mu \mp \Delta \mu\over D^{2}_{FW}}-g,
\label{en3}
\end{equation}
where $D_{FW}=y_1-y_2$ and we take into account that the
acceleration term depends on the coordinate $y$, e.g. [5]:
$g=a+a^2y$. For the average distance $Y=(y_1+y_2)/2$ to be at rest
$Y(\tau)=0$ the acceleration term $a$ must be balanced by the
gravity term ${\mu \over D^{2}_{FW}}$. We get
\begin{equation}
a={\mu \over D^{2}_{FW}}. \label{en4}
\end{equation}
Taking into account that in the lab frame the spacial coordinate
of four acceleration is related to $a$ as $a^x=\gamma a$ we have
\begin{equation}
\dot U^x=\gamma^3\dot V=\gamma a, \quad \dot V= {\mu \over
\gamma^2 D^{2}_{FW}}. \label{en5}
\end{equation}
It is clear that this coincides with equation (\ref{en2}).

The dynamical equation for the relative distance $D_{FW}$ directly
follows from (\ref{en3}) and (\ref{en4}):
\begin{equation}
\ddot D_{FW}=-{2\Delta \mu \over D_{FW}^{2}}-a^2D_{FW}=-{2\Delta
\mu \over D_{FW}^{2}}-{\mu^2\over D_{FW}^{3}}.
 \label{en6}
\end{equation}
It coincides with (\ref{en1}).

The last term on the right hand side of (\ref{en6}) is due to the
non-uniform acceleration force appearing in the Fermi-Walker
coordinates. Because it is $\propto \mu^{2}$, technically, it is a
post-Newtonian term. Since we consider the gravitational  force in
the Newtonian approximation in (\ref{en6}) it is important to
check whether or not post-Newtonian corrections to the
gravitational force are comparable with the acceleration term in
(\ref{en6}). In fact, as is described in standard handbooks, e.g.
[6], the post-Newtonian corrections are either proportional to
$\Delta \mu $ or $\dot y_{1,2}$. The mass difference and
velocities are assumed to be small and therefore, the terms in
(\ref{en6}) arising from the post-Newtonian corrections appear to
be small compared to the terms taken into account.

From (\ref{en6}) it follows that when the mass difference is
negative and $D_{FW}=2|\Delta \mu |$ the particles are at rest
with respect to each other. In this case the Fermi-Walker
coordinate system locally coincide with the Rindler one and the
particles accelerate indefinitely. Thus, unlike the Newtonian case
considered in the previous Section the particles accelerating
indefinitely being at rest with respect to each other must have
the small mass difference. This effect was first noted by Bondi
1957 [1]. It is obviously due to the non-uniform character of the
acceleration term.

\subsection{Solution of dynamical equations}

Since equation (\ref{e13}) contains only $V$ and $D$ it can be
used to express $V$ in terms of $D$
\begin{equation}
V={EP\over E_c^2+P^{2}}(1\mp {E_c\over EP}\sqrt{E_{c}^{2}-N^{2}}),
\label{e15}
\end{equation}
where $E_{c}$ is expressed through $D$ in equation (\ref{e11}),
$E$ and $P$ are given in equation (\ref{e12a}). As we discussed
above we assume that at the initial moment of time $t=\tau=0$ we
have $V=0$. That means that initially we have to choose the sign
$(-)$ in (\ref{e15}). However, under certain conditions discussed
below the direction of motion of the particles relative to each
other and, accordingly, $\dot D$, changes sign. At the turning
point $\dot D=0$ we have $N^{2}=E_{c}^2$. Since velocity $V$ must
grow monotonically according to (\ref{en2}) we must take the sign
$(+)$ in (\ref{e15}) after the turning point.

On the other hand equation (\ref{e14}) contains only $D$ and its
derivative with respect to the time $\tau$, and, therefore, it can
be integrated to obtain the dependence of $D$ on time. Explicitly
we have
\begin{equation}
\int^{D}_{D_{min}}{xdx \over \sqrt {R(x)}}=\tau/\mu, \label{e16}
\end{equation}
where
\begin{equation}
R(x)=(\mu^2+2\Delta \mu x)^{2}-N^{2}x^{2}.
\label{e17}
\end{equation}

The integral in (\ref{e16}) can be evaluated by a standard
substitution to give an explicit relation between $\tau $ and $D$.
However, the final expressions are rather cumbersome and we do not
show them here. Instead, in general, we analyse qualitatively
solutions to (\ref{e14}) based on analogy between this equation
and an equation describing a motion of a particle in a potential
well. For that we bring (\ref{e14}) to a standard form
\begin{equation}
{{\dot D}^{2}\over 2}+ U(D)={\cal E}, \quad U(D)=-{2\Delta \mu
\over D}-{\mu^{2}\over 2D^2},\label{e18}
\end{equation}
where
\begin{equation}
{\cal E}={4\Delta \mu^{2}-N^{2}\over 2\mu^{2}}={v_{in}^{2}\over
2}-{2\Delta \mu \over D_{in}}-{\mu^{2}\over 2D_{in}^2}.
\label{e19}
\end{equation}
Introducing natural units $\tilde U = {\mu^2\over \Delta \mu^{2}}
U$ and $\tilde D = {|\Delta \mu |\over \mu^2}D$ we can express
$\tilde U $ in terms of $\tilde D$ in a very simple form: $\tilde
U =\mp {2\over \tilde D}-{1\over 2{\tilde D}^2}$, where the sign
$-$ ($+$) corresponds to $\Delta \mu > 0$ ($\Delta \mu < 0$). The
dependence $\tilde U(\tilde D)$ is shown in Fig. 1.
\begin{figure}
\begin{center}
\vspace{1cm}
\includegraphics[width=14cm,angle=0]{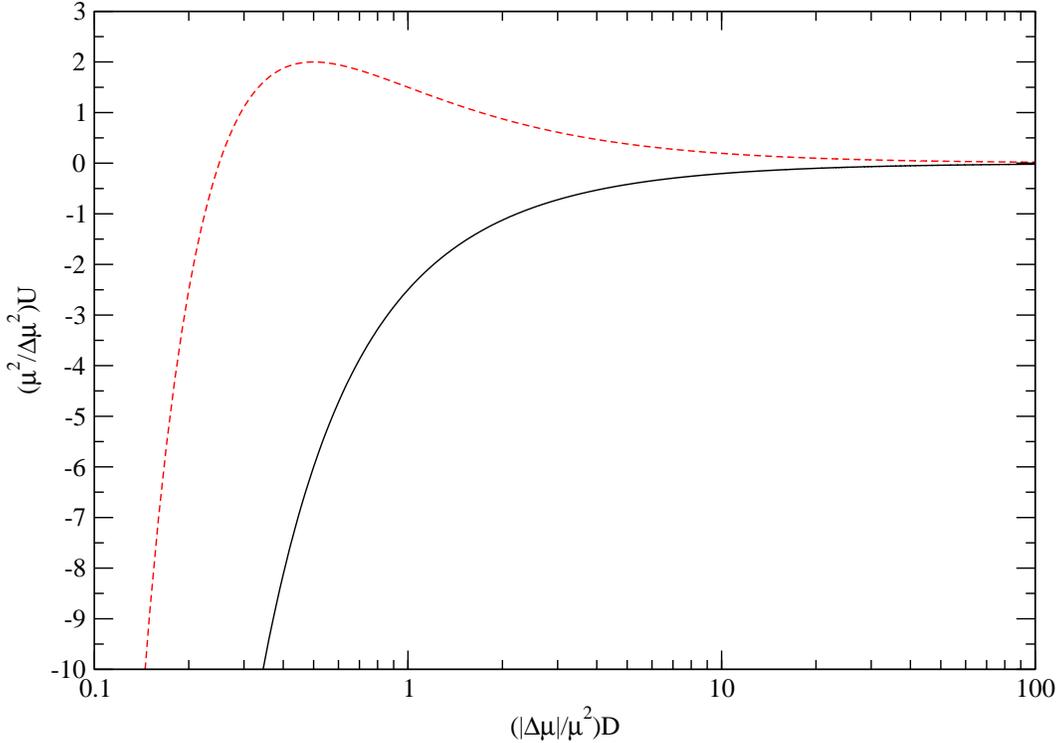}
\end{center}
\caption{The dependence of the potential $U$ on the spacial
coordinate $D$. The solid curve corresponds to the case $\Delta
\mu > 0 $ while the dashed one to the case $\Delta \mu < 0 $.}
\label{figg1}
\end{figure}

At first let us consider in detail an important case of zero norm
of the energy-momentum vector, $N^{2}=0$, and set, accordingly,
$P=E$.
 A simple analysis of equation
(\ref{e15}) shows that in this case there are no turning points,
the relative separation $D$ grows with time and the value of $V=1$
can be achieved in the asymptotic limit $\tau \rightarrow \infty$.
Therefore, in this case the system may accelerate indefinitely.

When $N^{2}=0$ equation (\ref{e15}) simplifies to
\begin{equation}
V={E^{2}-E_c|E_c|\over E^2+E_c^2}, \quad \gamma =
{(E^2+E_c|E_c|)\over 2EE_c} \label{e20}
\end{equation}
and from equation (\ref{e16}) we get
\begin{equation}
\tau={1\over 4\Delta \mu^2}(2\Delta \mu (D-D_{min})-\mu^{2}\log
({\mu^2 + 2\Delta \mu D\over \mu^2 +2\Delta \mu D_{min}})).
\label{e21}
\end{equation}

From equation (\ref{e20}) it follows that when $E_c > 0$ the
indefinite acceleration is possible only if $E_c \rightarrow 0$
when $\tau \rightarrow \infty$ and from the expression for $E_c$
(\ref{e11}) it is seen that the mass difference $\Delta \mu $ must
be negative for that. We consider below only this case in detail.
When $|\Delta \mu |\ne 0$ $E_c\rightarrow 0$ provided that
$D\rightarrow D_{crit}=\mu^2/(2|\Delta \mu |)$. Equation
(\ref{e21}) tells that the logarithm on the right hand side
diverges when $D\rightarrow D_{crit}$. That means that this limit
does correspond to the limit $\tau \rightarrow \infty$. Let us
estimate the dependence of the Lorentz factor $\gamma $ on time in
this case. To do so, we introduce a new variable $\Delta
=D_{crit}-D$ and substitute it to (\ref{e21}) assuming that it is
small. We get
\begin{equation}
\tau\approx {\mu^{2}\over 4\Delta \mu^2}\log ({\mu^{2}-2|\Delta
\mu |D_{in}\over 2|\Delta \mu |\Delta }),   \label{e22}
\end{equation}
and, substituting this result into equation (\ref{e20}) we have
\begin{equation}
\gamma \approx {\mu^2\over 4|\Delta \mu |D_{in}}\exp {4\Delta
\mu^2\over \mu^2}\tau.
 \label{e23}
\end{equation}
Equation (\ref{e23}) tells that when $D\approx D_{crit}$
acceleration is exponentially fast.

The degenerate case $\Delta \mu =0 $ must be analysed separately.
In this case from (\ref{e21}) we have
\begin{equation}
\tau={1\over 2\mu^2}(D^{2}-D_{min}^{2}), \label{e24}
\end{equation}
and the distance $D$ increases indefinitely with time. From
equation (\ref{e20}) we obtain
\begin{equation}
\gamma \approx {\mu \over 2D_{min}}\sqrt {2\tau }. \label{e25}
\end{equation}

Now let us turn to the general case $N^{2} \ne 0$. Setting $\dot
D=0$ in (\ref{e18}) we get a general equation for the turning
points
\begin{equation}
D_{1,2}={\Delta \mu \over {\cal E}}(-1\pm \sqrt{1-{\mu^{2}{\cal
E}\over 2\Delta \mu^2}})={\Delta \mu \over {\cal E}}(-1\pm
{\sqrt{N^2}\over 2\Delta \mu }). \label{e26}
\end{equation}
Equation (\ref{e26}) tells that the turning points exist only when
$N^{2} > 0$. Their number depends on signs of ${\cal E}$ and
$\Delta \mu $. When $\Delta \mu > 0$ the potential $U(D)$ is
negative, see Fig. 1, and therefore, the relative motion is finite
for ${\cal E} < 0$ with one turning point \footnote{Let us
remember that we consider only positive values of $D$.}
\begin{equation}
D_{1}={\Delta \mu \over |{\cal E}|}(1+{N\over 2\Delta \mu}).
\label{en26}
\end{equation}
In the opposite case ${\cal E} > 0$, and, accordingly, $N <
2\Delta \mu $, the motion is unbound and the relative distance $D$
grows indefinitely with time.

When $\Delta \mu < 0$ the potential $U(D)$ acquires positive
values for $D > {\mu^{2}\over 4|\Delta \mu |}$, see Fig. 1. It
tends to zero when $D \rightarrow \infty $ and has a maximum at
$D=D_{crit}$. Note that from the condition $U(D_{crit})={\cal
E}={2\Delta \mu^2\over \mu^2}$ we get $N^{2}=0$ there. The
character of the relative motion depends on whether ${\cal E}$ is
negative, belongs to the interval $0 < {\cal E} < {2\Delta
\mu^2\over \mu^2}$ corresponding to $0 < N < 2|\Delta \mu |$, or
${\cal E} > {2\Delta \mu^2\over \mu^2}$ and, accordingly, $N^{2} <
0$. When the energy ${\cal E}$ is negative the motion is bound
with one turning point
\begin{equation}
D_{1}={|\Delta \mu |\over |{\cal E}|}(-1+ {N\over 2\Delta
\mu}).\label{e27}
\end{equation}
In the intermediate region $0 < {\cal E} < {2\Delta \mu^2\over
\mu^2}$ there are two turning points
\begin{equation}
D_{\pm}={|\Delta \mu | \over {\cal E}}(1\pm {N\over 2\Delta
\mu}).\label{en27}
\end{equation}
When $D_{in} < D_{-}$ the motion is bound while for $D_{in} >
D_{+}$ $D$ grows indefinitely. Finally, when $N^{2} < 0$ the
motion is always unbound.

When the motion is bound the velocity $\dot D$ changes sign after
the turning point and in this case we should use the sign $(+)$ in
(\ref{e15}). Taking into account that $D$ is decreasing after the
turning point and that $E_{c} \propto D^{-1}$ we see from
(\ref{e15}) that the velocity $V\rightarrow 1$. The particles tend
to collide. However, our assumption that $D \gg \mu $ breaks down
in this case and we cannot describe the motion at scales $D\sim
\mu $ within the framework of our formalism. Note that we consider
in this study only pairs of particles with strictly zero angular
momentum. In the situation when the particles have a small but
nonzero angular moment they would miss each other and after
certain moment of time the distance $D$ would become negative. In
this case the analysis of this paper can be repeated without any
major change for negative values of $D$ and one would conclude
that for such parameters of motion there is another symmetric
turning point at negative values of $D$. Thus, the relative motion
of a pair of particles with small but nonzero orbital momentum
would be periodic much similar to the case of ordinary particles
with positive masses.

Now let us consider the case of the unbound motion and estimate
the maximal value of the Lorentz gamma factor the system can
reach. As follows from our previous discussion when $N^2\ne 0$ the
distance $D$ grows indefinitely. That means that the energy in the
comoving frame, $E_{c}$, must tend asymptotically to $2\Delta
\mu$. Note that when $\Delta \mu < 0$ the asymptotic value of
$E_c$ is negative. We have from (\ref{e15}) setting $E_{c}=2\Delta
\mu $ there
\begin{equation}
V={1\over 4\Delta \mu^2+E^{2}-N^2}(E\sqrt{E^2-N^2}-2\Delta \mu
\sqrt {4\Delta \mu^2 -N^2}).
\label{e28}
\end{equation}
Equation (\ref{e28}) tells that when $\Delta \mu > 0$ the last
term in the brackets is negative and the asymptotic value of
velocity is smaller than 1. Large values of $V$ can be achieved in
the opposite  case $\Delta \mu < 0$ assuming $|N^2| \ll \Delta
\mu^2 $. In this case we expand expressions in  (\ref{e28}) in the
Taylor series in $|N^2|/\Delta \mu^2 $ to obtain
\begin{equation}
V=1-{N^{4}\over 32\Delta \mu^2 E^2}, \label{e29}
\end{equation}
and, accordingly
\begin{equation}
\gamma\approx {1\over \sqrt{2(1-V)}}={4|\Delta \mu E|\over |N^2|}.
\label{e30}
\end{equation}
Equation (\ref{e30}) tells that for fixed values of $E$ and
$\Delta \mu < 0$ the gamma factor can be made arbitrary large by
choosing arbitrary small values of $|N^{2}|$. This conclusion is
in agreement with our previous finding that the system accelerates
indefinitely when $N^2=0$.

\section{Methodological comments}

Here I would like to make comments on several methodological
issues related to the problem.

1) At first glance the fact that the 'average' position of the
pair $(x_1+x_2)/2$ always grows with time may seem to be in
contradiction with the law of conservation of the centre of mass
of the system. This contradiction is resolved by observation that
for the system containing particles of opposite masses position of
the centre of mass, $R$, is determined by a {\it difference } of
positions of particular particles. Say, in the Newtonian
approximation we have $R=m_1x_1+m_2x_2=(\mu+\Delta
\mu)x_1-(\mu-\Delta \mu)x_2$. In the relativistic case the
situation is analogous for systems with $N^2 > 0$. In the opposite
case the notion of centre of masses is ill defined. Indeed,
introducing the velocity of a coordinate system, where the centre
of mass is at rest in a standard way  as $V_{cm}=P/E$ [6] we see
that when $N^{2}=0$ $V_{cm}=1$ and when $N^{2} < 0 $ $V_{cm}$
formally exceeds the speed of light. It is obvious that the notion
of the centre of mass is redundant in both cases.

2) In Introduction of their paper the authors of [3] claim that
the conception of PMT put forward by the author of this note is
related to the problem of indefinite acceleration of two
gravitationally interacting particles. This statement needs, in my
opinion, a clarification. First, let me note that as it is
discussed above even in the case when only a finite acceleration
of the particles is attained  PMT is still possible in a situation
where production of such pairs is provided by some physical
mechanism. Second, the conception of PMT, in general, does not
rely on gravitation interactions at all. In particular, in paper
[2] I consider a model where there is a continuous flow of
positive energy from some spacial regions having negative energy
to other regions with positive energy provided by hydrodynamical
effects. In this model the space-time is assumed to be flat and
gravitational interactions are absent. Moreover, in order to
construct a PMT it is not necessary to invoke objects having
negative rest masses. It is enough to consider a medium with
positive comoving energy density violating the weak energy
condition [2]. Additionally, there are ways of constructing PMT,
where gravitational interaction plays a totally different role,
say, transferring the energy from a non-stationary system having
negative mass to gravitational waves, as for example in the model
of a rotating relativistic string connected by two negative mass
monopoles [2], [7]. The effects related to the dynamics of free
negative mass particles are clearly irrelevant to such systems.

3) The authors of [3] claim that it is impossible to obtain, in
principal, an indefinite acceleration of the system containing two
particles with opposite signs of masses. One may think that this
clearly contradicts to the Bondi's result [1] and the results of
this paper. The conundrum is resolved by observation that the
authors of [3] consider only {\it relative} motions while Bondi's
analysis as well the analysis in this note also deal with the
motion of the pair of particles as a whole with respect to an
inertial frame.

\section{Conclusions}
In this note we show by elementary means that in the weak limit
approximation a pair of particles having opposite values of masses
can be accelerated indefinitely provided that  the energy-momentum
vector characterising the system is null. The system can also be
accelerated to arbitrary large Lorentz factors when the mass
difference $\Delta \mu < 0$ and the norm of the energy-momentum
vector is sufficiently small.

Assuming that there is a process of production of such pairs and
that the positive mass particles are intercepted with a target
while the negative mass particles fly away it is possible to
transfer to the target any desired amount of energy. In a more
natural situation one can also consider a theory where the
positive and negative mass particles interact differently with a
normal matter. A general situation of this kind where there is a
persistent transfer of energy from a subsystem having negative or
almost zero energy (like this pair ) to a subsystem with positive
energy was dubbed by us 'Perpetuum Motion of Third Kind' (PMT)
[2]. Note, however, that it is just a classical analog of the well
known instability of a quantum system with a number of negative
energy states unbound from below.

The question of whether the existence of PMT or ever accelerating
pairs of particles is a paradox depends, in our opinion, on
definition of what paradox is. On one hand, for example, Bonnor
1989 states 'I regard the runaway (or self-accelerating ) motion
... so preposterous that I prefer to rule it out by supposing that
inertial mass is all positive or all negative' [8]. Clearly,
existence of PMT can also be classified as a kind of runaway. On
the other hand, no laws of physics are broken in such systems. We
believe that the existence of runaways of these kinds is dangerous
for theories where they present. To exemplify, an indefinite
concentration of energy of different signs in spatially separated
regions could lead to a highly inhomogeneous space-time hardly
compatible with presence of any life. Therefore, such theories
should be ruled out though some additional study of them in
General Relativity may be of some interest.

Since in our approximation only linear metric perturbations and
one next-to-the-leading order term determined by the acceleration
of the pair as a whole are taken into account, it is interesting
to estimate what kind of corrections can be obtained by
considering other higher order terms quadratic in metric
perturbations? For a non-relativistic motion with $V \ll c$ for
this purpose one can use the well known Einstein-Infeld-Hoffmann
equations of motion (e.g. [6]). In this way it is convenient to
consider particles with a large mass difference as well as systems
with non-zero angular momentum. There are, however, many
corrections, which are absent in such a treatment, notably the
emission of gravitational waves. Therefore, a self-consistent
relativistic treatment of the problem in the next to the weak
field approximation must be based on the second-order formalism of
Havas and Goldberg 1962. Such an approach is left for a possible
future work.

Although in this paper we consider only particles with no internal
structure our analysis may also be valid for a pair of extended
objects with total energies of opposite signs provided that they
have a sufficiently large separation distance and that their
relative velocities are sufficiently small. For example,  Deser
and Pirani [9] considered the behaviour of systems with all
possible inertial/gravitational mass signs and noted that a pair
of geons having opposite signs of their total energies would
behave as a pair of point particles in the corresponding limit.

It is also interesting to point out that the notion of 'Perpetuum
Motion of Third Kind' was introduced in the context of
thermodynamical systems having negative temperatures, where one
can withdraw heat from a negative temperature reservoir and
convert it completely to work, see e. g. [10], p. 176. Since
thermodynamical systems with negatives masses of their components
should have negative temperatures (e.g. [11]) there is a link
between thermodynamical properties of such systems and the ones
discussed in this paper. In particular, a runaway process
occurring in a thermodynamical system having two subsystems
containing particles with opposite signs of masses has been
discussed in e.g. [12]. It has been mentioned that this process is
analogous to the self-acceleration of a pair of particles with
opposites signs of their masses.

\begin{acknowledgments}
I am grateful to S. Deser, J. F. Gonza'lez Herna'ndez,  I. D.
Novikov, K. A. Postnov and V. N. Strokov  for useful comments.

This work was supported in part by the Dynasty Foundation, in part
by ''Research and Research/Teaching staff of Innovative Russia''
for 2009 - 2013 years (State Contract No. P 1336 on 2 September
2009) and by RFBR grant 11-02-00244-a.

\end{acknowledgments}


\begin{thebibliography}{}

\bibitem{1} Bondi, H., Reviews of Modern Physics, 29, 423 (1957)


\bibitem{2} Ivanov, P., Physics Letters B,  680, 212 (2009)

\bibitem{3} Shatskii, A. A., Novikov, I. D., Kardashev, N. S., Phys. Usp. 54, 399 (2011)


\bibitem{4} Havas, P., Goldberg, J. N., Physical Review, 128, 398
(1962)

\bibitem{5} Misner, C. W., Thorne, K. S., Wheeler, J. A.
'Gravitation', San Francisco: W. H. Freeman, (1973)


\bibitem{6} Landau, L. D., Lifshitz, E. M.,
'The Classical Theory of Fields, Fourth Edition: Volume 2 (Course
of Theoretical Physics Series)', Butterworth-Heinemann; 4 edition
(1980)

\bibitem{7} Martin, X., Vilenkin, A., Phys. Rev. D 55, 6054 (1997)

\bibitem{8}  Bonnor, W. B., General Relativity and Gravitation, 21, 1143 (1989)

\bibitem{9} Deser, S., Pirani, F. A. E., Annals of Physics, 43, 436 (1967)

\bibitem{10}  Landsberg, P. T., 'Thermodynamics and Statistical Mechanics', Courier Dover Publications
(1987)

\bibitem{11} Vysin, V., Physics Letters, 2, 32 (1962)

\bibitem{12} Pollard, D., Dunning-Davies, J., Il Nuovo Cimento,
110B, 857 (1995)




\end{thebibliography}
\end{document}